\documentclass{flexibility}

\usepackage{times}  
\usepackage{epsfig}
\usepackage[TABBOTCAP]{subfigure}
\usepackage{tabularx}
\usepackage{graphicx} 
\usepackage{color}
\usepackage{xspace}
\usepackage{thumbpdf}
\usepackage{listings}
\usepackage{verbatim}
\usepackage{hyperref}
\usepackage{booktabs}
\usepackage{colortbl}

\hypersetup{pdfstartview=FitH,pdfpagelayout=SinglePage}

\conferenceinfo{HotNets 2015} {}
\CopyrightYear{2015}
\crdata{X}
\date{}

\usepackage{xcolor}

\makeatletter
\def\@copyrightspace{\relax}
\makeatother

\begin{document}

%
\conferenceinfo{WOODSTOCK}{'97 El Paso, Texas USA}

\title{Flexibility of Networks: Towards a new measure for analyzing the network design space}
\title{Flexibility of Networks: Towards a new measure for network design space analysis?}
\title{Flexibility of Networks: a new measure for \\ network design space analysis?}

\numberofauthors{1} 
%
\author{
%
%
\alignauthor
Wolfgang Kellerer, Arsany Basta, Andreas Blenk\\
       \affaddr{Chair of Communication Networks\\ Department of Electrical and Computer Engineering}\\
       \affaddr{Technical University of Munich}\\
       \email{\{wolfgang.kellerer,arsany.basta,andreas.blenk\}@tum.de}
}
\additionalauthors{Additional authors: John Smith (The Th{\o}rv{\"a}ld Group,
email: {\texttt{jsmith@affiliation.org}}) and Julius P.~Kumquat
(The Kumquat Consortium, email: {\texttt{jpkumquat@consortium.net}}).}
\date{30 July 1999}

\maketitle
\begin{abstract}
Flexibility is often claimed as a competitive advantage when proposing new network designs. However, most proposals provide only qualitative arguments for their improved support of flexibility. Quantitative arguments vary a lot among different proposals.
A general understanding for flexibility is not yet clearly defined, leaving it to the reader to draw the right conclusions based on background information. 
The term flexibility is commonly defined as the ability to adapt to changes. 
For networks, flexibility would refer to the ability to adapt the available network resources, such as flows or topology, to changes of design requirements, e.g., shorter latency budgets or different traffic distributions. 
Recent concepts such as Software Defined Networking, Network Virtualization and Network Function Virtualization have emerged claiming to provide more flexibility in networks. 
Nevertheless, a deeper understanding of what flexibility means and how it could be quantified to compare different network designs remains open.
In this paper, we ask whether flexibility can be a new measure for network design space analysis. 
As it is quite challenging to formulate a flexibility measure that covers all network aspects, we propose an initial set of flexibility aspects to start grounding guidelines.
Our initial selection is backed up by an analysis of Software Defined Networking, Network Virtualization and Network Function Virtualization for their support of the selected flexibility aspects. 
Our research methodology is based on a systematic approach that leads to network design guidelines with respect to flexibility.
\end{abstract}

\section{Introduction}
Flexibility has become a key design objective for networks and respective proposed control and data plane mechanism today. 
For example, more than one third of the publications presented at ACM SIGCOMM in 2014 mention flexibility in their description.

In fact, heterogeneous requirements from different application domains demand for networks to be designed for flexibility. 
These requirements include the ability to add new flows or even virtual networks on demand without influencing existing flows/networks, the ability to temporarily extend a network topology to serve events, and the ability to reconfigure the network in real-time for resilience, e.g., for industrial applications, 
to give some examples.

Let us take a popular sports event as another example~\cite{Erman2013}. For a short period of time thousands of users demand network resources to send videos or to obtain additional information about the game. To meet those demands, network resources have to be allocated in the best possible way to scale with the increased number of users. Topology might be adapted to allow multicasting of extra information. However, this is only needed for some hours. 
A network that can satisfy these requirements is commonly said to be flexible.

Flexibility can be defined in different domains and from different viewpoints. 
For networks, which is the focus of this paper, flexibility refers to the ability of a network to adapt its resources such as flows or topology to changes of requirements.
This adaptation to changes may include the adaptation of the network configuration, the network topology or the network functions and their placement.

Note that in this paper we focus on network aspects of flexibility. In general, in a communication system, flexibility may also refer to aspects of software implementation, operating systems, protocol stack design, application design, etc., which are out of scope for our discussion here. 


In the recent years, a number of technologies have emerged claiming to provide flexibility in networks.
One widely accepted approach is the concept of Software Defined Networking (SDN)~\cite{McKeown2008} separating the data plane from a logically centralized control plane with a standardized interface allowing programmability and hence flexibility in networks. 
SDN-based network control can be complemented by the concept of Network Virtualization (NV)~\cite{Chowdhury2008} where network resources can be operated on logical, hence virtual level on a physical network substrate. 
The concept of virtualization has also been extended to network functions. 
Network Function Virtualization (NFV)~\cite{nfvmano} allows to provide network functions such as gateways and middleboxes in software and to run them on commodity hardware, e.g., in data centers. 

Although new technologies evolved that increase the ability of a network to be adapted, a clear definition of what flexibility means for networks is missing. 
Moreover, there is no common agreement on a quality indicator quantifying a network's flexibilty. 
Such quality indicator could be defined similar to what has been defined for Quality of Service (QoS). QoS has been introduced  to provide a common understanding about network support for service level performance \textit{aspects}, in particular, data rate, delay and jitter.

For flexibility, we raise the question whether it will be a new measure for network design space analysis.
Moreover, we advocate to come up with a network flexibility measure, e.g., "Quality of Flexibility" (QoF), describing a common set of flexibility \textit{aspects}.
Similar to QoS, where the importance of aspects such as data rate and delay varies among different service requirements, flexibility depends on the requirements as well.
For some network scenario the placement of functions may be important, for another its scale in topology size.
Hence, we are not aiming at quantifying flexibility of networks as a singular comparative metric, but rather through a set of flexibility aspects.
To be able to quantitatively compare different network designs with respect to their flexibility, a common definition of main flexibility aspects is indispensable. 
\begin{figure}[t]
	\centering
	\includegraphics[trim=0.5cm 0.5cm 0.5cm 0.5cm, clip=true, width=0.35\textwidth]{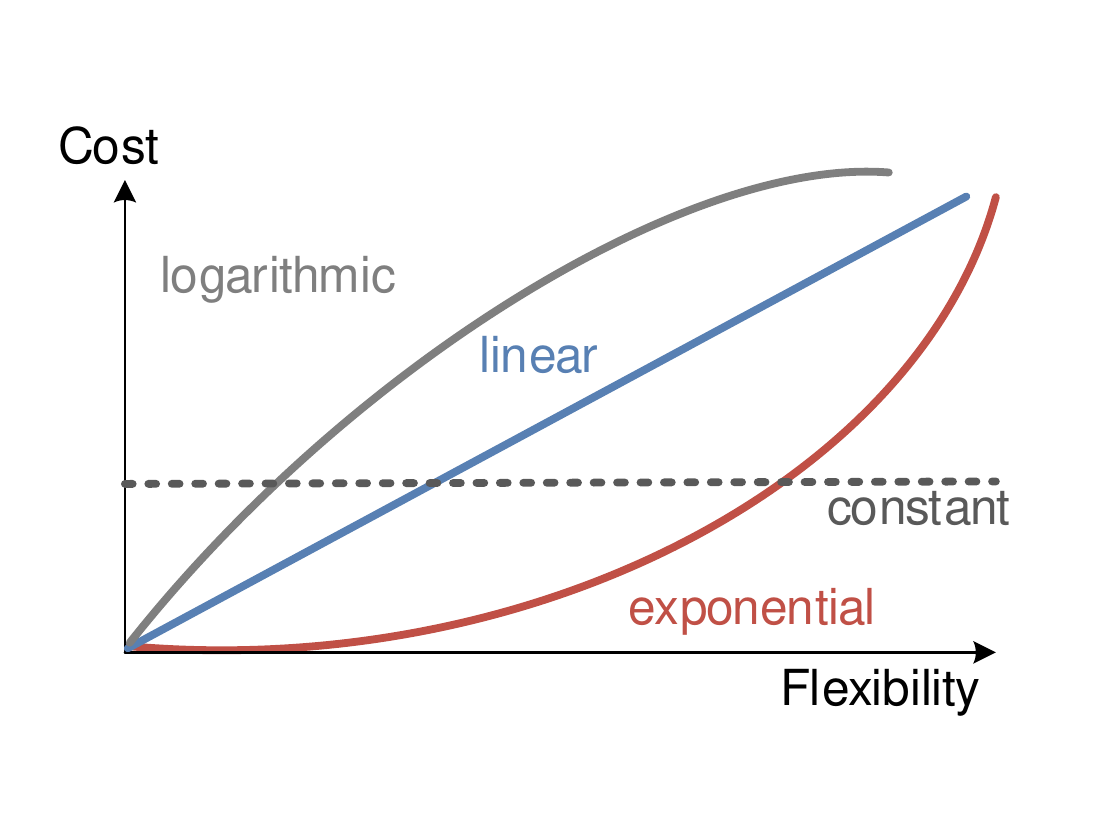}
	\vspace{-10px}
	\caption{Trade-off between flexibility and cost}
	\label{fig:flex_vs_cost}
\vspace{-10px}
\end{figure}

The cost incurred to provide flexibility is important to consider when we compare different network designs with respect to network flexibility. 
Cost involved in providing flexibility may include, e.g., additional network load for signaling, additional data path latency, and the number of reconfigurations needed to change a network state.

Whereas ongoing research is focusing to come up with new network designs and protocols to realize the SDN concept or NFV for various use cases, a fundamental analysis of the cost-benefit trade-off for flexibility in networks is missing. 
Figure~\ref{fig:flex_vs_cost} illustrates how such trade-off analysis could look like. A common understanding might be that cost rises with increased flexibility, e.g., signaling overhead increases with the number of supported configuration parameters. However, we lack quantitative trade-off evaluation results for different network design choices.  
Would costs rise linearly, logarithmic or exponential with increased flexibility?
Or would the costs rather remain constant?
For a quantitative analysis and, in particular, for the comparison of design choices, we need a define the aspects that comprise a network flexibility measure.

The main contributions of this paper are (a) to show that flexibility is used for network evaluation in various ways, but a common definition of a flexibility measure is missing, (b) to propose an initial set of flexibility aspects for a common network flexibility measure, (c) to advocate that in network research a deeper analysis of the fundamental design space of networks for flexibility with respect to cost is needed, and (d) to layout guidelines on the methodology and approach that we intend to follow to tackle this fundamental problem.

In the remainder of this paper, we analyze state of the art approaches for their support of flexibility in Section 2. In Section 3, we propose a selection of flexibility aspects as a basis for a common network flexibility measure. In Section 4, we discuss how our proposal applies to emerging technologies such as SDN, NV and NFV based on examples. In Section 5, we introduce a methodology framework that aims at defining a measure for network flexibility.

\section{Flexibility in Literature}\label{sec:literature}
In this section, we analyze the state of the art that argues about flexibility. 
We extract the definition of flexibility applied in each targeted use case and, if possible, show how flexibility is expressed via quantitative measures. 

\subsection{Flexible Network Architectures}
Anderson et al.~\cite{Anderson2005} discuss the flexibility gain of network virtualization. They argue that virtualization is needed in order to provide flexible experimentation with traffic from the current Internet. Furthermore, they introduce two views of a future architecture, the purist view and the pluralist view. As the architecture remains in place a long time, the purists aim for architectural flexibility. This means that the architecture should only provide mechanisms to be changed over large time scales. In contrast, the pluralists want to provide the ability to add or augment overlay networks when needed. They argue that flexible adding or removing overlays, i.e., changing virtual topologies, provides the needed flexibility. 

Greenberg et al.~\cite{Greenberg2009} propose VL2, a scalable and flexible data center network. They argue that data centers should provide an adaptive and flexible resource allocation in order to achieve a high resource efficiency. Although the authors do not define flexibility as the dominant system attribute, they mention agility as the key property. In the data center context, the key to achieve high utilization ''is the property of agility - the capacity to assign any server to any service.''

The authors of~\cite{Mukerjee2013} investigate the flexibility of inserting new technologies in existing infrastructures. 
They define ''Flexibility'' as ''the ability for'' an ''approach to adapt to changes in topology over time (...) as well as failures''.
They quantify the flexibility for different technologies. 
For instance, flexibility (fault tolerance) vs. achieved throughput. While one approach is more flexible (failure resilient), it adds overhead, thus, decreases throughput.

\subsection{Flexible Mobile Networks}
Jin et al~\cite{Jin2013a} tackle the challenges of the cellular core network. 
They say that the current mobile core network is ''inflexible'' for three reasons: they ''forward all traffic through the P-GWs'', ''P-GWs are not modular'', carriers cannot ''mix and match capabilities from different vendors (e.g., use a firewall from one vendor, and a transcoder from another)''. 
They propose the scalable architecture SoftCell that can make fine-grained policies for the mobile core network devices. SoftCell uses so called flexible, high-level service policies. Operators can use these policies to redirect traffic through middleboxes, which are operated according to the demands of subscribers. The high level policies are realized via switches that are deployed close to the base stations. The core switches enable forwarding to the needed middleboxes, i.e., network functions. 

\subsection{Flexible Network Management}
Arumaithurai et al.~\cite{Arumaithurai2014} propose Function-Centric Service Chaining (FCSC). FCSC is based on Information Centric Networking in order to make the management of networks that use virtualization for dynamic function placement more flexible. They see flexibility as the ability to adapt faster to failures and to change middleboxes more quickly. 
More in detail, ''an efficient service chaining network should support (...) changes in a flexible way - (...) middleboxes should be able to determine the functions of a flow themselves and the changes should take effect immediately.''

\subsection{Flexible Data Plane}

In order to make current switches more flexible in terms of QoS, Sivaraman et al.~\cite{Sivaraman2013a} propose to add an FPGA-based extension to switches. This extension provides more capabilities to control the fast-path and queuing behavior of switches. 
In particular, ''the data plane should be flexible enough to handle diverse and unanticipated application requirements.''
It is also argued that such ''flexibility could be realized easily in software router running on a general-purpose microprocessor.'', i.e., software routers, but that these lack providing the same performance as hardware implementations, or software solutions extended with hardware.

Similar to the previous concept, Hwang et al.~\cite{Hwang2014b} say that software solutions running on commodity servers, whose hardware is extensively exploited via software extensions, e.g. DPDK, provides ''far greater flexibility'' than existing purpose-built hardware. 
They propose NetVM, which ''enable(s) in-network services'', e.g., firewalls or proxies, ''to be flexible created, chained and load balanced.'' 
For instance, they propose a shared switching memory inside hypervisors to avoid memory migration during the migration of virtual machines.


\subsection{Flexible Protocols}
Han et al.~\cite{Han2013} are focusing on solutions for the congestion problem of networks. They see router-assisted congestion control algorithms as not flexible enough as the end-point is dependent on the feedback from the network. On the other hand, they see pure end-point based solutions as not as efficient as router-assisted solutions. Accordingly, they present a framework called FCP. FCP relies on both, i.e., it provides flexible end-point realization that can incorporate congestion feedback from the network. Thus, it provides the ''flexibility to ensure that new behaviors can be implemented to accommodate potential changes in communication patterns.''

\subsection{Flexible Traffic Control}
Chowdhurry et al.~\cite{Chowdhury2013} propose Sinbad that lets applications make decisions about where to steer their file traffic. Thus, Sinbad avoids congested network links by avoiding network traffic hotspots. It increase the flexibility of cluster file systems by adapting the replica destinations.

Vissicchio et al.~\cite{Vissicchio2014, Vissicchio2015} introduce Fibbing, an architecture that ''readily supports flexible load balancing, traffic engineering, and backup routes''. Fibbing provides a way to have a control plane that runs physically distributed but is still centrally controlled. For this, they introduce fake nodes and links in order to indirectly impact the path calculation of the distributed control plane. In this, the advantages of both worlds should be combined. The authors also mention that ''while more flexible (e.g., enabling stateful control logic) than Fibbing, SDN requires updating the switch-level rules one-by-one'' thus ''forgoes the scalability and reliability benefits of distributed routing.''

\hspace{5mm}

In summary, we can observe that flexibility is a key requirement for network design in the related work. However, several different perspectives of flexibility are taken and a common understanding of network flexibility as measure is missing so far. Nevertheless, we can observe common aspects of network flexibility among the different publications. In the following, we are going to extract those to come up with a set of flexibility aspects as part of a common measure.






\section{Towards a Flexibility Measure}\label{sec:metrics}

Network flexibility can be expressed with respect to many different network parameters, e.g., set of possible configurations or number of locations for function placement. 
There is no unified measure that can express how flexible a network is, i.e., to quantify flexibility for comparing network design choices. 
There is also a lack of quantitative analysis for the incurred cost, i.e., overhead, resulting from increasing network flexibility in a network. 

Defining a measure for network flexibility is not a trivial problem as a lot of the involved network parameters are depending on each other. 
The main reason for this dependency is the huge variety of the parameters. 
For example, the flexibility of migrating a virtual network depends on the migration mechanism, the size of the network, the topology of the network, the hypervisor used, the physical technology, the function to be migrated, etc.
For a flexibility measure, the challenge is to quantify the resulting values of each parameter according to their dependencies. 

Network flexibility as a measure is mostly used with a specific objective in mind, e.g. "network A can be re-configured faster than network B", focusing on a selected, narrowed down set of parameters, which we could already observe in Section 2.
Hence, in order to come up with a common measure, the challenge is to find reasonably independent flexibility aspects combining some of the parameters to support an intuitive understanding of flexibility.

In the following, we provide an initial set of such flexibility aspects.
For each, we list the parameters defining the aspect as well as the cost involved for achieving the respective flexibility.

A network can be assessed in terms of its flexibility to change its \textbf{configuration}. 
The configuration can either be a single parameter or a state change, i.e., "re-configuration" or rather an addition to the possible set of configurations that the network supports. 
Network configuration flexibility can be assessed in terms of the number or set of possible configurations, where a larger set of possible configurations adds to flexibility. 
Another assessment of configuration flexibility is time, in which a configuration is either changed, added or enforced. 
Network configuration flexibility can be expressed for flow configuration, function configuration and parameter configuration.

Another aspect of network flexibility is the ability to change the deployment location of network functions within a given network, i.e., network function \textbf{placement}. 
Network function placement allows to meet different latency requirements and also the combination of functions, i.e., chaining.

A third area for network flexibility is the aspect of \textbf{scale}. This includes the ability of a network to scale its resources, e.g., to add link capacity, or to scale the allocated resources to network flows, e.g., allocate more capacity to a flow. It also refers to the ability to scale and apply adaptations to the network topology, e.g., scale the network size through adding nodes and links or change the network connectivity from a tree to a mesh topology. 

Note that the proposed aspects can be considered as lenient examples for an initial set of network design choices in the context of flexibility. These flexibility aspects can be extended through new networking concepts, technologies or future design requirements.


\subsection{Flow Configuration} 

Flow configuration describes the course of flows inside a network through configuring forwarding policy for a flow on each network hop. 
Flow configuration can be considered as an elementary attribute for configuration flexibility. Having the ability to change the configuration of the flow policies offers traffic steering, which in turn can bring more flexibility to networks. Such flexibility can be related to the magnitude and granularity of flow configurations, more possible configurations would reflect to higher flexibility. As an example, a network element that can support only forwarding of packets is less flexible than an element that can provide both forwarding and duplicating packets on multiple ports for instance. 
Flexibility of flow configuration can be also coupled with the time required to change such configuration. Network elements can vary from not being able to change the flow configuration on run-time, i.e., static, to elements that can support run-time flow configuration.  

It is very important to note that there is a cost to support higher flexibility in terms of flow configuration. From an operational aspect, changing the configuration of network nodes requires additional control, which might impose latency and data overhead. 
From a performance perspective, changing the flow configuration "steering" might lead to service interruption or even to network instability. 

\textbf{parameters:} e.g., set of flow configurations, support for run-time configuration

\textbf{cost:} e.g., control latency, control overhead, network stability, flow interruption 


\subsection{Function Configuration} 
Function configuration denotes the ability of configuring the functionality of  network elements 
such as firewalls, NATs, proxies, load balancers, etc. 
Nowadays, programmable switches are being introduced which allow the operator to change and tweak their network function. Hence, a programmable resp. configurable network element can be another driver to increase network flexibility. 
Flexibility of function configuration can be assessed in terms of the set of possible functions supported by the programmable network element.
The run-time support to change the function configuration can be considered as another main enabler for higher flexibility.    
The cost of flexible function configuration can be observed in terms of latency or control overhead. It may also require additional resources or capabilities on the programmable network elements compared to conventional network elements. Function configuration might also impact the performance of the data plane. For instance, if the function configuration feature is only supported by software implementations that run on the general-purpose computing of a network element, a relative decline in performance could be observed compared to other functions that are implemented and integrated in the hardware. 


\textbf{parameters:} e.g., set of function configurations, support for run-time configuration

\textbf{cost:} e.g., function configuration latency, resource overhead, control overhead, data plane performance, data processing latency

\subsection{Parameter Configuration} 
In addition to flow configuration, which describes the data flow in a network, and function configuration, that denotes the functionality in a network, there is a third type of configuration which is parameter configuration. 
Parameter configuration concerns changing the values and policies to be used by each network function. This means that flow path and the network functionality remain the same, however the parameters configured on those functions can vary. For example, for a priority queuing scheduler, parameter configuration would be setting the priority values for the receptive queues, or it would be the maximum rate for a port shaper. 
More possible parameter configurations and the ability of changing the network parameters on run-time would imply more flexibility. Similar to flow and function configuration, there is a cost induced by parameter configuration coming from control and data plane performance.   

\textbf{parameters:} e.g., set of parameter configurations, support for run-time configuration

\textbf{cost:} e.g., parameter configuration latency, resources overhead, control overhead, data performance

\subsection{Function Placement} 

The placement of a function within a network defines the possible locations for network functions. The function placement has a direct impact on the network performance, e.g., the SDN controller placement with respect to switches and its impact on control latency. 
Dynamic placement adds an additional dimension to flexibility in case changing the function placement is supported through, e.g., migration techniques for virtual functions. 

The placement flexibility is directly influenced by the set of possible locations to place a function. More potential locations have the degree of freedom to place network functions such that diverse or even more strict requirements can be satisfied. The connectivity given between the set of location can also play a role in the overall flexibility. A dynamic function placement that can change on run-time offers more flexibility than a static placement.

\textbf{parameters:} e.g., set of potential locations, connectivity between locations, static or dynamic placement

\textbf{cost:} e.g., control and data latency, control and data throughput, state consistency, synchronization overhead (depending on migration mechanism), interruption during migration (depending on migration time) 

\subsection{Resource Allocation} 

The allocation of resources denotes the flexibility of a network to change the assignment of network resources to flows or functions. It is decided based on the possible resources, e.g., network element CPU or link capacity, that can be allocated to individual flows or functions. For example, for a network element that has two functions which share equally its resources, e.g., CPU or memory, resource allocation flexibility would mean that we can assign 80\% of the resources to one of the functions.
Higher resource allocation flexibility would be achieved with more possible types of resources that can be assigned. Flexibility is also related to the granularity of such resource assignment. 
Adding more resource allocation flexibility in network elements means more complexity and management overhead. It could also mean that part of the resources can be utilized by the manager that enforces the resource assignment.

\textbf{parameters:} e.g., granularity of assignment, set of possible resources to be assigned

\textbf{cost:} e.g., network element management overhead, network element complexity, resources overhead

\subsection{Topology Adaptation} 

The adaption of network topology describes the flexibility of a network to change its topology structure through adding or removing nodes or links. Topology adaptation flexibility can be reflected by the network technology. As an example, adding a node to an optical topology can require more effort (in terms of tuning and setup) compared to adding a node in an IP topology. The network technology can also refer to physical compared to logical topologies. A logical topology in this sense has more flexibility to adapt its mapping on the network infrastructure, while a physical network is restricted by its set of physical nodes and links. Additionally, discovery protocols also play a role to support the flexibility in adapting a network topology. A topology that runs an automated discovery protocol, which provide on run-time topology adaptation, is more flexible than a topology that has to be manually configured.  
Flexibility of topology adaptation comes with a cost in terms of additional protocols and management overhead. The topology adaptation protocols might also have a cost in terms of network resources, i.e., additional resources needed to run these protocols. 



\textbf{parameters:} e.g., technology, discovery protocols, run-time adaptation

\textbf{cost:} e.g., topology adaption latency, resource overhead, signaling and management overhead, protocol complexity, data throughput and latency

\begin{center}
\begin{table*}[ht]
	\centering
    \caption{Concepts vs. Flexibility Aspects. (\checkmark): main target, (-): out of scope, (+): provides support}
\label{tab:concepts_flexibility}
    \begin{tabular}{| l |l |l |l | l |  l | p{2cm} | }
    \hline
Concept & Flow Config & Function Config & Parameter Config & Function Placement & Resource Allocation & Topology Adaptation \\ \hline \hline
SDN & \checkmark & -- & -- & + & +  & + \\ \hline
NV & -- & -- & + & \checkmark & \checkmark & \checkmark \\ \hline
NFV & -- & \checkmark & + & \checkmark & \checkmark & -- \\ \hline
    \end{tabular}
\end{table*}
\end{center}

\section{How flexibile are recent \\ networking concepts?}

Emerging technology concepts to provide flexibility in networks include SDN, NV and NFV. 
In order to illustrate how a quantitative analysis of the flexibility vs.\ cost trade-off could look like, we describe and discuss selected use cases in the following. 
We have applied our selected flexibility aspects (Section 3) to examples from the state of the art  of each of the three concepts.
Table~\ref{tab:concepts_flexibility} illustrates which flexibility aspects are supported by each of them. 


\subsection{Software Defined Networking}

SDN was developed to target programmable flows and to centralize network control, which contributes to flexibility in terms of flow configuration. This flexibility needs to be assessed in terms of the number of possible configurations. OpenFlow (OF)~\cite{McKeown2008}, 
which is  the most commonly used protocol to implement SDN, has an upper boundary in its flexibility due to the limited set of configurations specified in the specification of each OF protocol version. 
In addition to flexibility of flow configuration, SDN's network control can also indirectly support the flexibility of network functions placement~\cite{liu2014}, flow configuration~\cite{jain2013b4} and topology adaptation~\cite{berde2014}. 

An example for the trade-off between SDN's flow configuration flexibility and its cost can be observed in~\cite{levin2012}. 
This work illustrates the cost of state synchronization between distributed SDN controllers with the application of load balancing. 
The evaluation looks at two controllers that exchange link utilization information towards two servers. 
The target is to apply load balancing among the two servers by consistent flow configuration based on the exchanged state. 
It is shown that more frequent state synchronization, which translates into signaling and processing overhead, is needed to achieve the targeted load balancing. 
As we can observe, a concrete flexibility vs. cost trade-off provided by SDN does not come for granted, but might induce cost on network operation.   





\subsection{Network Virtualization}

NV abstracts network resources from physical infrastructure with the scope of adding flexibility to network resources.
With existing networking hardware, virtualization can contribute to flexibility in terms of flow and topology adaptation, flexibility of function placement through migration of virtual nodes as well as flexibility of parameter configuration for the abstract virtual resources. 

Addressing migration for instance to evaluate the flexibility of virtual networks,~\cite{keller2012} shows a study for live migration solution of virtual switches. A live migration provides a high flexibility in adapting the virtual network topology. The evaluation shows that the introduced solution can successfully achieve migration without packet loss, i.e., transparent to the service. However, an extra software layer is added that comes at a control overhead of 7\%. This means that gains in terms of topology flexibility offered by NV might have drawbacks on performance, i.e., cost. 



\subsection{Network Function Virtualization}

NFV leverages virtualization to functionality, where functions get developed as software and are executed on commodity hardware.     
Having programmable hardware can offer flexibility to define and program function configuration. NFV can also provide flexibility in terms of flow scale by being independent from networking hardware, e.g., scale up resources assigned to a network function or scale out a function on multiple hardware entities. Software functions, wich are independent from hardware, also contribute to the function placement flexibility.

An example to show the trade-off between the flexibility and cost of NFV can be seen in~\cite{sherry2012}. This work investigates the opportunity to virtualize network middleboxes and to convert them into software functions that run in a cloud. Middleboxes contribute to a large fraction of network domains, e.g., enterprise, thus software inter-changeable middleboxes can promise a huge increase in flexibility. The evaluation shows that flexibility of software middleboxes can induce a cost in terms of increased latency depending on the cloud provider and solution taken. It is also shown that the cost of traffic overhead with software middleboxes can be up to an additional 52\% in the worst case. 

\hspace{5mm}

Overall, we can observe that our proposed network flexibility aspects well apply for latest technology concepts.
First quantitative cost analysis is provided, however, as a common measure is missing, a general quantitative analysis and comparison of different design choices remains challenging. Moreover, most current work is highlighting improvements with respect to flexibility. An analysis of network flexibility limits is missing as well. Such should be part of a comprehensive analyis of the network design space to be able to show why a design choice is flexible and to what extent. One should always be careful when reading evaluation statements just claiming improved flexibility as such.


\section{Framework towards quanti-\\fying Flexibility} \label{sec:framework}
\begin{figure}[t]
	\centering
	\includegraphics[width=0.42\textwidth]{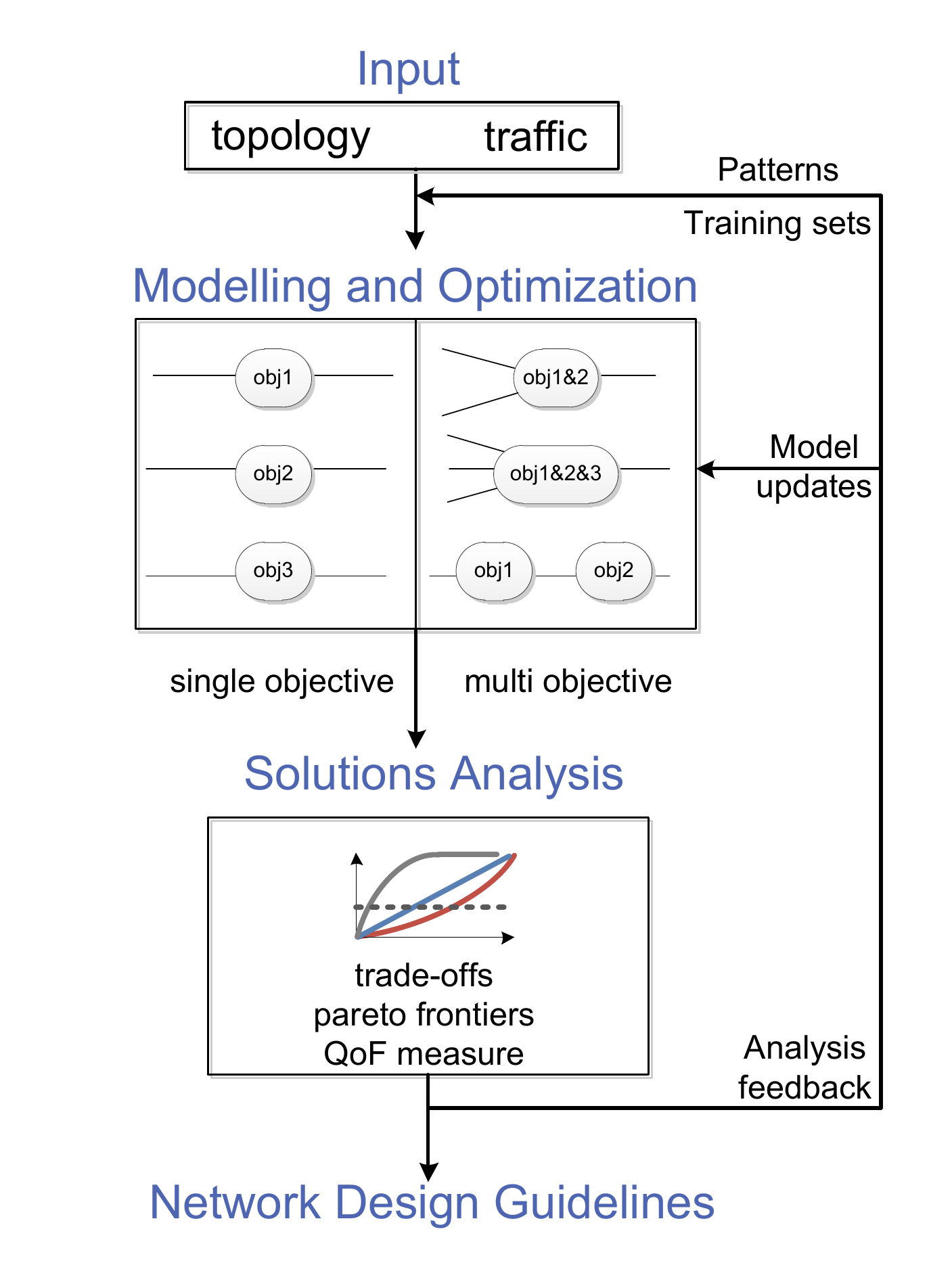}
	\caption{Methodology framework towards network flexibility measure. Based on an initial solution analysis, observed patterns and new training sets are fed back for modeling and optimization. The models are updated based on insights from consecutive solution analysis}
	\label{fig:framework}
\end{figure}

In this section, we outline our methodology and approach towards a measure for network flexibility. We define a systematic approach that would lead to guidelines on how to design a network with respect to flexibility. The framework consists of three main building blocks, namely objective definition, solution analysis and guideline formulation, as shown in Figure~\ref{fig:framework}. 


\subsection{Input}
As we define the flexibility of a network in terms of its ability to adapt, the input of a flexibility framework should use input that shows changing behavior in time and space.
This assumption is based on the fact that, e.g., network users are changing their locations over time.
Such diurnal patterns are regularly observed in current traffic measurements taken from different types of networks.
In order to support spatial traffic patterns, i.e., network traffic occurs with varying intensity for different locations, also spatial behaviors need to be considered when analyzing flexibility.
Beside varying traffic patterns, the underlying network topology may strongly impact the performance of network architectures. Thus, different types of topologies serve as input for flexibility analysis.

\subsection{Modeling and Optimization}

This first step is considered as the core of the whole framework. The selected flexibility aspects, mentioned above in Section~\ref{sec:metrics}, are modeled as network design optimization problems. The optimization problems target specific technology aspects and concept details, e.g., model the control and data plane split in SDN networks or model the logical mapping of virtual networks on the physical substrates. The objective definition then incorporates the different network requirements that can be inferred from today's networks, e.g., minimize data plane delay, minimize the management overhead by considering re-configurations, or maximize the support for drastic traffic changes or fluctuations. The resulting network design based on the defined objectives would be influenced by the input, which can be narrowed down on an abstract level to two main contributors, namely, different network topologies and differing traffic distributions. Our approach aims at altering the optimization's topology and traffic input. This might result in different network design solutions, which might show trade-offs or Pareto frontiers. 

\subsection{Solutions Analysis}

The next step is the analysis of the solutions. This brings us closer to defining guidelines for a flexibility measure. In the first step, the optimization problems are solved with varying input of network topology and traffic distributions resulting in a whole set of solutions. The analysis of these solutions derives patterns from the solutions and moves a step forward towards the flexibility measure.

One example would be to deduce which network design could support more variations of the topology and traffic input, hence, infer higher flexibility. We could consider the number of supported variations as a flexibility measure.
Besides, initial solution analysis might reveal so called problem-solution patterns. A pattern might be that one architecture fits to topologies showing a specific characteristic, such as high betweenness centrality, while another architecture might fit best to sparsely connected network topologies.
Furthermore, the solution analysis could also reveal that for a set of traffic patterns and topologies, a flexibility parameter does not need to be considered at all. For instance flexibility in configuration might not be needed as all available technologies might not support the time dimension of the topology or traffic input. 


\subsection{Formulation of Network Design Guidelines}
The last step in our framework is to incorporate all information about the optimization objective, used input, and results of the solution analysis to be able to formulate guidelines for a flexible network design. 
This step requires linking and combining the different aspects involved in the challenge to define a quantitative flexibility measure.

The outcome of the solution analysis could then be used as feedback for model updates and also to thin out and substantiate the input data for the modeling and optimization step.
Thinning out and substantiating the input data should lead to experiments in which conditions are investigated which actually affects the variation of the optimization results.
This cycle could be repeated several times within the proposed framework till it converges to a set of clear and solid guidelines.

This modeling/optimization/analysis cycle needs a clear stopping condition.
As an example, a simple stopping condition might be that the whole process stops if the complete problem and solution space was exhaustively analyzed, i.e., all possible combinations of parameters and input data were investigated.
However, such an analysis might be infeasible due to the size of the problem and solution space. Thus, sophisticated mechanisms for identifying when an analysis truly converges, i.e., does not produce false positive conclusions, need to be established.

\section{Conclusion}
Flexibility is commonly used as a differentiating feature in recent proposals for network designs. 
However, quantitative arguments are often missing in order to express clearly which flexibility aspects are addressed to which extent and what costs are incurred to state that one network design is more flexible than another design. 
Therefore, we advocate that in network research a deeper analysis of the fundamental design space of networks for flexibility with respect to cost is needed.
Moreover, we ask the question whether network flexibility is constituting a new measure for network design space analysis. 
We claim that with emerging networking concepts such as SDN, NFV and NV, network flexibility will most likely become a new measure in network research and development in the future.
In our initial proposal such flexibility measure is not a single parameter but includes several flexibility aspects including the ability to adopt dynamic changes in network configuration, the ability to place network functions and the ability to scale the network topology in size.
Our selection of flexibility aspects is backed up by an analysis of SDN, NV and NFV based on concrete use cases. The latter show that an initial evaluation of network flexibility with respect to costs is already taking place that could benefit from our proposed measure. Accordingly, we proposed an initial framework for investigating flexibility aspects in a well-defined manner.
The framework consists of multiple steps that should be repeated iteratively, finally leading to clear and solid design guidelines for network architectures, which support flexibility in the paper's context.
Benefits include to reveal  limits of flexibility and to be able to compare among different approaches.


\section*{ACKNOWLEDGMENT}
This work is part of a project that has received funding from the European Research Council (ERC) under the European Union's Horizon 2020 research and innovation program (grant agreement No 647158 - FlexNets). This work reflects only the authors' view and the funding agency is not responsible for any use that may be made of the information it contains.
\bibliographystyle{unsrt}
\bibliography{rw}  
%
%
\end{document}